\documentstyle[11pt]{article}
\begin{document}
\begin{center}
\begin{Large}
{\bf Time and Temperature Dependent Correlation
Functions of $1D$ Models of  Quantum Statistical Mechanics}
\end{Large}

\vspace{36pt}
Vladimir Korepin\raisebox{2mm}{{\scriptsize $\dagger$}}
and Nikita Slavnov\raisebox{2mm}{{\scriptsize $\ddagger$}}

\vspace{20pt}

~\raisebox{2mm}{{\scriptsize $\dagger$}}
{\it ITP, SUNY at Stony Brook, NY 11794-3840, USA.}\\
korepin@insti.physics.sunysb.edu

\vspace{6pt}

~\raisebox{2mm}{{\scriptsize $\ddagger$}}
{\it Steklov Mathematical Institute,
Gubkina 8, Moscow 117966, Russia.}\\
nslavnov@class.mi.ras.ru
\vspace{72pt}
\end{center}
{\small
We consider gapless models of statistical mechanics. At zero 
temperatures correlation functions decay asymptotically as powers of 
distance , however at non-zero temperatures these
 correlations decay exponentially. We used an example of a solvable model
to  find the formula, which describes long distance and large time
asymptotic  of correlation function of local fields. The formula describes 
the correlation function of local fields  at any temperature and arbitrary 
 coupling constant.

\vspace{50pt}

PACS numbers: 05.30.-d, 05-50.+q, 65.50.+m, 67.40.Db}

\newpage

We consider quantum models of statistical mechanics in one space and 
one time dimension. We shall study models with one type of excitation
(particle). 

First let us discuss the zero temperature case. The ground states of these
 models, in which we are interested are Fermi spheres. All possible states
 of the particle are filled from $-k_F$ to $k_F$. We shall use $k$ for 
the momentum of the particle and $\varepsilon$ for the energy. The Fermi
velocity $v_F$ is defined as:
\begin{equation}\label{velocity}
v_F=\left.\frac{\partial\varepsilon}{\partial k}
\right|_{k=k_F}.
\end{equation}
At large distance $x$ the  correlation function of local quantum fields
$\psi(x,t)$ decays as a power law:
\begin{equation}\label{asym1}
\langle\psi(0,0)\psi^\dagger(x,t)\rangle
\begin{array}{c}
{}\\
\longrightarrow\\
\mbox{\raisebox{2mm}{$x\to\infty\atop t\to\infty$}}
\end{array}
\frac{1}{(x-v_Ft)^{2\Delta^+}(x+v_Ft)^{2\Delta^-}}.
\end{equation}
The powers $\Delta^\pm$  can be called conformal dimensions \cite{BPZ}.
At very small temperatures $T$, the Fermi velocity still can be defined, 
and the  correlation function decays exponentially
\begin{equation}\label{asym2}
\langle\psi(0,0)\psi^\dagger(x,t)\rangle_T
\begin{array}{c}
{}\\
\longrightarrow\\
\mbox{\raisebox{2mm}{$x\to\infty\atop t\to\infty$}}
\end{array}
\exp\left\{-\frac{2\Delta^+\pi T}{v_F}
|x-v_Ft|-\frac{2\Delta^-\pi T}{v_F}|x+v_Ft|\right\}.  
\end{equation} 

Increasing the temperature will destroy the Fermi surface and the
 temperature 
dependence of the correlation function will become more complicated. 
In this paper we have found a universal formula (\ref{asym3}) which 
describes the asymptotic correlations at any temperature. We consider 
the example of the Bose gas with delta interactions. Our description of the 
model  follows Chapters I and XVIII of the book \cite{KBI}.

In order to define the model let us consider many-body quantum mechanics
 of $N$ identical particles. The Hamiltonian of the Bose gas can be
 represented as
\begin{equation}\label{Ham1}
{\cal H}_N=-\sum_{j=1}^{N}\frac{\partial^2}{\partial x_j^2}
+2c\sum_{N\ge j>k\ge1}\delta(x_j-x_k).
\end{equation}
The coupling constant $c$ is positive. The model also can be reformulated 
as a model of quantum field theory, in which it is  described by the 
canonical quantum Bose fields $\psi(x,t)$. The commutation relations are 
standard 
\begin{equation}\label{commut}
\begin{array}{rcl}
{\displaystyle [\psi(x,t),\psi^{\dagger}(y,t)]}&=&
{\displaystyle \delta(x-y),}\\
\vspace{-0.5mm}&\vspace{-0.5mm}&\vspace{-0.5mm}\\
{\displaystyle [\psi(x,t), \psi(y,t)]}&=&{\displaystyle 0.}
\end{array}
\end{equation}
The Hamiltonian of the same model can now be written as an operator 
in the Bosonic Fock space:
\begin{equation}\label{Ham2}
{H}=\int_{-\infty}^{\infty} dx
\left({\partial_x}\psi^{\dagger}(x)
{\partial_x} \psi(x)+
c\psi^{\dagger}(x)\psi^{\dagger}(x)\psi(x)\psi(x)\right).
\end{equation}
In this form the model is known as the Quantum Nonlinear Schr\"odinger
equation.

The Bethe Ansatz for the model was discovered by E.~H.~Lieb and W.~Liniger
\cite{LL}. The physics of interacting bosons ($c>0$) is different from 
that of free bosons ($c=0$). For interacting bosons the Pauli principle
is valid in momentum space \cite{IK} and the  zero temperature ground 
state is a Fermi sphere. In order to describe the ground state exactly 
it is more convenient to use the spectral parameter $\lambda$ instead of 
the physical momentum of the particle $k$. The spectral parameter $\lambda$ 
coincides with the momentum only when the density of the gas vanishes.
 At positive  density, the  momentum $k$ can be expressed as a function of
 spectral parameter
(see (\ref{mom0})).The ground state can be described by the Lieb--Liniger
integral equation:
\begin{equation}\label{rho0}
\rho(\lambda)-\frac{1}{2\pi}\int_{-q}^{q} K(\lambda,\mu)
\rho(\mu)\,d\mu=\frac{1}{2\pi}.
\end{equation}
Here $q$ is the value of the spectral parameter on the Fermi surface, and
\begin{equation}\label{kernel}
K(\lambda,\mu)=\frac{2c}{c^2+(\lambda-\mu)^2}.
\end{equation}
The density function $\rho(\lambda)$ shows the distribution of particles 
in the Fermi sphere: $L\rho(\lambda)\,d\lambda$ gives the number of 
particles in the interval $[\lambda,\lambda+d\lambda]$ and  $L$  denotes
the length of the box. The excitation energy $\varepsilon(\lambda)$ (energy
of the particle) can be described by a similar integral equation
\begin{equation}\label{eps0}
\varepsilon(\lambda)-\frac{1}{2\pi}\int_{-q}^{q} K(\lambda,\mu)
\varepsilon(\mu)\,d\mu=\lambda^2-h.
\end{equation}
It vanishes on the Fermi surface $\varepsilon(\pm q)=0$. Here h is the
 chemical potential. We consider the case $ h > 0$ .

The momentum of the particle can be expressed as a function of the spectral 
parameter:
\begin{equation}\label{mom0}
k(\lambda)=\lambda+\int_{-q}^{q} \theta(\lambda-\mu)
\rho(\mu)\,d\mu.
\end{equation}
Here
\begin{equation}\label{theta}
\theta(\lambda)=i\ln\left(\frac{ic+\lambda}{ic-\lambda}\right).
\end{equation}
The Fermi velocity is equal to
\begin{equation}\label{velocity1}
v_F=\left.\frac{\partial\varepsilon(\lambda)}
{\partial k(\lambda)}
\right|_{\lambda=q}=\frac{\varepsilon'(q)}{k'(q)}.
\end{equation}
Where the denominator can be expressed in terms of $\rho$ as
\begin{equation}\label{denom}
\frac{\partial k(\lambda)}{\partial\lambda}=2\pi \rho(\lambda).
\end{equation}
This derivative on the Fermi surface is called the dressed charge
\begin{equation}\label{charge}
Z=2\pi\rho(q).
\end{equation}
It defines the conformal dimensions $\Delta^\pm$ from (\ref{asym1})
\begin{equation}\label{confdim}
2\Delta^+=2\Delta^-= \frac{1}{4Z^2}.
\end{equation}
The Luttinger liquid derivation of these formul\ae~was first found by
F.~D.~M. Haldane \cite{H}. 

At very small temperatures the formula (\ref{asym2}) describes 
correlations.  We calculated the correlation function at arbitrary 
temperature. Before formulating our results let us describe 
the thermodynamics.

At positive temperatures it is necessary to introduce two density 
functions $\rho_p(\lambda)$ and $\rho_t(\lambda)$. The number of 
particles in the interval $[\lambda,\lambda+d\lambda]$ is equal to
$L\rho_p(\lambda)\,d\lambda$. The joint number of particles and holes in
$[\lambda,\lambda+d\lambda]$ is equal to $L\rho_t(\lambda)\,d\lambda$.
C.~N.~Yang and C.~P.~Yang discovered  \cite{YY} a system of equations 
which describe the state of thermodynamic equilibrium
\begin{equation}\label{thermodynamics} \begin{array}{l}
{\displaystyle 2\pi\rho_t(\lambda)}={\displaystyle 1+
\int_{-\infty}^{\infty} K(\lambda,\mu)
\rho_p(\mu)\,d\mu,}\\
\vspace{-0.5mm}\\
{\displaystyle\frac{\rho_p(\lambda)}{\rho_t(\lambda)}}=
{\displaystyle\frac 1{1+e^{\frac{\varepsilon(\lambda)}T}},}\\
\vspace{-0.5mm}\\
{\displaystyle\varepsilon(\lambda)}={\displaystyle\lambda^2-h- 
\frac{T}{2\pi}\int_{-\infty}^{\infty} K(\lambda,\mu)
\ln\left(1+e^{-\frac
{\varepsilon(\mu)}T}\right)\,d\mu.}
\end{array}
\end{equation}
 The existence 
of stable excitation (particle) also was discovered in \cite{YY}. 
This is a consequence of integrability of the model. The energy of this 
excitation is $\varepsilon(\lambda)$ and the momentum is
\begin{equation}\label{momT}
k(\lambda)=\lambda+\int_{-\infty}^{\infty} \theta(\lambda-\mu)
\rho_p(\mu)\,d\mu.
\end{equation}
The velocity of the excitation is
\begin{equation}\label{velocity2}
v(\lambda)=\frac{\partial\varepsilon(\lambda)}
{\partial k(\lambda)}
=\frac{\varepsilon'(\lambda)}{k'(\lambda)}.
\end{equation}
The denominator in (\ref{velocity2}) is
\begin{equation}\label{denom1}
\frac{\partial k(\lambda)}{\partial\lambda}=2\pi \rho_t(\lambda).
\end{equation}

The correlation function at positive temperature is defined as
\begin{equation}\label{corr1}
\langle\psi(0,0)\psi^\dagger(x,t)\rangle_T=
\frac{\mathop{\rm tr}\nolimits
\left(e^{-\frac{H}{T}}\psi(0,0)\psi^\dagger(x,t)\right)}
{\mathop{\rm tr}\nolimits\left(e^{-\frac{H}{T}}\right)}.
\end{equation}
Here the trace is taken over the  space of all states. We evaluated the long 
distance and
large time  asymptotic of this correlation function
\begin{equation}\label{asym3}
\langle\psi(0,0)\psi^\dagger(x,t)\rangle_T
\begin{array}{c}
{}\\
\longrightarrow\\
\mbox{\raisebox{2mm}{$x\to\infty\atop t\to\infty$}}
\end{array}
\exp\left\{\frac{1}{2\pi}\int_{-\infty}^{\infty}
\frac{d\lambda}{2\pi\rho_t(\lambda)}
|x-v(\lambda)t|\ln\left|\frac
{
e^
{
\frac{\varepsilon(\lambda)}{T}
}-1
}
{
e^
{
\frac{\varepsilon(\lambda)}{T}
}+1
}
\right|\right\}.
\end{equation}
This is our main result.

The key point of our derivation is the representation of the correlation 
function as a determinant of Fredholm integral operator \cite{KKS1}.
On the basis of this representation, a system of integrable equations 
was derived \cite{KKS2}. These equations can be solved
by the Riemann--Hilbert problem technique  and the long distance  asymptotic 
can be evaluated. This method of evaluation of correlation functions 
is described in detail in Parts III and IV of the book \cite{KBI}.

\section*{Acknowledgments}
This work was supported by NSF Grant PHY-9321165,
RFBR Grant  96-01-00344 and INTAS-01-166-ext.

\end{document}